\documentstyle[preprint,aps]{revtex}
\def\complex{\kern.1em{\raise.47ex\hbox{ $\scriptscriptstyle |$}}\kern-.40em{\rm
C}}
\title{Geometric Forces on Point Fluxes\\ in Quantum Hall Fluids}
\author{J.~E.~Avron}
\address{ Department of Physics, Technion, 32000 Haifa, Israel}
\author{ P.~G.~Zograf}
\address{Steklov Mathematical Institute, 191011
St.-Petersburg, Russia} 
\begin{document}
\tightenlines
\draft
\maketitle
\begin{abstract}
\noindent We study the forces that act on a point flux carrying an integral
number of flux units in quantum Hall fluids. Forces due to external fields,
Lorentz and Magnus type forces, and the forces due to mutual interaction of
point fluxes are considered. The forces are related to the adiabatic curvature
associated with families of Landau Hamiltonians. The problem displays distinct
features of the quantum Hall fluids with point fluxes on the plane and on the
torus, which, however, agree at the ``thermodynamic'' limit. \end{abstract}
\pacs {PACS numbers : 73.40.Hm, 74.60.Ge}
\narrowtext

A {\em point flux} in two dimensions is the magnetic analog of a point charge. It
is created by an (infenitesimally thin) Aharonov-Bohm flux tube that
orthogonally pierces the two dimensional surface in question. For reasons that
shall become clear later, we restrict ourselves to cases where the flux $\Phi$
is an integer multiple of the quantum flux unit $\Phi_0=\frac{h c}{e}$. 

A point flux in vacuum does not interact directly with electric or magnetic
fields, nor with other point fluxes (by the linearity of Maxwell equations).
However, point fluxes associated with, say, a two dimensional quantum Hall fluid
can interact through the electrons that fill the Landau level. 

There are several forces that we consider: the force due to an external electric
field, the force due to the point flux motion and the mutual forces that a
collection of point fluxes apply one on another. It turns out that the forces
depend, among other things, on the topology of the two dimensional manifold in
question: the forces that act in the plane and on the torus are different. In
both cases the forces are geometric in character and are related to the
adiabatic curvature \cite{berry,simon83}. 

The problem we study is related to, but distinct from, the study of the forces
that act on vortices in superfluids and superconductors
\cite{tinkham,volovik,ant}. The gauge field of a point flux is formally
identical to the velocity field of a vortex (the flux is an analog of the
vorticity), and in both cases the forces that arise have geometric
interpretation \cite{ant}. However, the physics is different and so are the
results.

There are also relations to the fractional quantum Hall effect
\cite{laughlin,stone,hlr} and to anyon physics \cite{lf,wilczek} where weakly
interacting composite objects made up of a charged particle attached to a point
flux, play a central role. 

Consider a family $H(A)$ of Landau Hamiltonians in two dimensions associated
with gauge fields $A(\phi,p_1,\dots, p_N)$ depending on $N+1$ complex parameters
. The parameter $p_j\in\complex\, (j=1,\dots, N)$ gives the position of the
$j$-th point flux on the two dimensional surface. The parameter
$\phi\in\complex$ is associated with a (constant in space) gauge field that
creates an external electric field,
$E=E_1+iE_2$, in two dimensions. More precisely, the Landau Hamiltonian $H(A)$
is given formally by
\begin{equation} H(A)=\frac{2\hbar^2}{m}\, D_A^* D_A,\quad\quad
D_A=\frac{\partial}{\partial\bar z}-iA_{\bar z}, \label{landau}\end{equation}
where $A=A_zdz+A_{\bar z}d\bar z$. The gauge field $A$ splits into three parts:
\begin{equation} A(\phi,p_1,\dots, p_N)= A^0+A^\phi + \sum_{j=1}^N \Phi_j\,
A^{p_j};
\label{aA}\end{equation}
$\Phi_j$ is an integer multiple of the unit of quantum flux $\Phi_0$ (in our
units $\Phi_0=2\pi$). The gauge field \begin{equation} A^0
=-\frac{1}{2}By(dz+d\bar z) \label{B}\end{equation} gives rise to the constant
magnetic field
$B$. When we consider such field on a torus of area $L^2$, we assume that $B$
satisfies Dirac's quantization condition, that is, $BL^2/2\pi$ must be an
integer. There is no constraint on $B$ for the Euclidean plane. The gauge field
\begin{equation} A^\phi =
\frac{1}{2}(\bar\phi\,dz+\phi\,d\bar z), \label{C}\end{equation} with $\phi$ a
function of time, is associated with a constant (in space) electric field
\begin{equation} E= -\frac{1}{2c}(\dot{\bar\phi}\,dz+\dot\phi\,d\bar z),
\label{E}\end{equation} where $\dot\phi$ is the time derivative of $\phi$. 

The gauge field $A^p$ is associated to the unit point flux located at $p$. It is
a solution of
$dA^p=2\pi\,\delta(z-p)\,dx\wedge dy,$ where $z=x+iy$. In the Euclidean plane a
canonical solution is the rotational invariant point flux: \begin{equation}A^p=
-\frac{i}{2}\,\left(\frac{dz}{z-p}-\frac{d\bar z}{\bar z-\bar p}\right).
\label{pfp}\end{equation} This solution is characterized by $A^p_z$ having a
simple pole at $p$ with residue $-i/2$. In the case of the (unit) torus, the
canonical gauge field associated to a point flux turns out to be
\begin{equation}A^p_z=-\frac{i}{2}\,
\frac{\vartheta'}{\vartheta}\Big(z-p-\frac{1}{2}-\frac{i}{2}\Big)
-\pi\Big(\mbox{Im}\,p+\frac{1}{2}\Big),\label{pft}\end{equation} where the theta
function is defined, as usual, by
\begin{equation}\vartheta(z)=\sum_{n=-\infty}^\infty e^{-\pi n^2}e^{2i\pi
nz}.\label{theta}\end{equation} When $\phi$ and $p_j$ depend adiabatically on
time, i.e. we consider $\phi(t/\tau)$ and $p_j(t/\tau)$ in the limit
$\tau\to\infty$, the electric field $E=O(1/\tau )$ is weak and the velocities
$\dot p_j =O(1/\tau )$ are small. We are interested in the forces on the point
flux to the same order in $\tau$. In particular, we do not calculate any forces
that are of order $1/\tau^2$ (e.g. forces that are proportional to, say, $E\dot
p$).

Our main result is as follows. On the torus represented by the square of size
$L\times L$ with opposite sides identified, the quantum expectation value for
the force acting on the $j$-th point flux in a full Landau level is
\begin{equation} \langle F_j\rangle=\Phi_j\,\frac{e^2}{\hbar c}\, \left(E -
\frac{i}{c L^2}\sum_{k=j}^N\,
\Phi_k\,\dot p_k\right)+O(1/\tau^2).\label{main}\end{equation} The expression in
brackets in the right hand side of Eq.~(\ref{main}) has physical significance:
the motion of the point fluxes creates an emf. With $N$ moving point fluxes the
total emf,
$V=V_1+iV_2$, about the parallel and meridian of the torus, as we shall explain,
is
\begin{equation} V={L}\,\left(E - \frac{i}{c L^2}\, \sum_{k=1}^N \ \Phi_k \,
\dot p_k\right).\label{totalemf} \end{equation} Eq.~(\ref{main}) can therefore
be re-written as \begin{equation} \langle F_j\rangle=\Phi_j\,\frac{e^2}{\hbar
c}\, \frac{V}{L}+O(1/\tau^2),\end{equation} where $\frac{e^2}{\hbar c}$ is the
fine structure constant. Since $\frac{e^2}{\hbar c}\,\Phi= e\,\frac{\Phi\,
e}{\hbar c}= e\,\frac{\Phi }{\Phi_0}$, Eq.~(\ref{main}) may be interpreted as
saying that a point flux interacts with an {\em effective electric field} $V/L$
{\em as if} it were {\em electrically charged} with a charge that is a multiple
of the electron charge by the number of flux quanta. 

In the thermodynamic limit
$L\to\infty$, the effective electric field $V/L$ becomes the external electric
field $E$, by Eq.~(\ref{totalemf}). In this limit, or, in other words, in the
Euclidean plane, Eq.~(\ref{main}) reduces to \begin{equation} \langle F_j\rangle=
\frac{e^2}{\hbar c}\,\Phi_j E+O(1/\tau^2).\label{forceplane} \end{equation}
There is no Lorentz or Magnus type force on a point flux in the plane, and point
fluxes do not mutually interact. This is, of course, quite unlike vortices in
superfluids and superconductors which do interact with each other \cite{tinkham}
and experience a Magnus force \cite{volovik,ant}. We see that the forces on
point fluxes in the plane, $L=\infty$, and the torus, $L<\infty$, have distinct
character. The term with $k=j$ in the sum in Eq.~(\ref{main}) is a Lorentz (or
Magnus) type force which is self-interacting being proportional to $\Phi_j^2$.
The sum over $k\neq j$ describes a velocity dependent mutual interaction of
point fluxes that violates Newton's third law. 

We shall give an outline of a formal derivation of the above results. Let us
start with Eq.~(\ref{pft}) which gives the gauge field of a point flux on torus.
In analogy with the gauge field of a point flux in the plane, it is
characterized by a holomorphic function of $z$, with a simple pole at $p$ of
residue $-i/2$ which is doubly periodic in $p$. For the sake of simplicity let
us assume that the torus $T$ is given by the factor of $\complex$ modulo the
unit square lattice (the case of a square lattice of arbitrary size may be
reduced to that one by scaling). The holomorphic function $\vartheta (z)$,
Eq.~(\ref{theta}), has exactly one simple zero per lattice cell located at the
points $(1/2+m,\ i/2+in)$ with integer $m,n$, and has the following
transformation properties with respect to translations by 1 and $i$:
\begin{equation}\vartheta(z+1)=\vartheta(z),\quad \vartheta(z+i)=e^{\pi-2i\pi
z}\,\vartheta(z).\end{equation} Put $\zeta=p+1/2+i/2$ and consider the function
\begin{equation} f(z;\zeta)=e^{-2\pi iz\,\mbox{\scriptsize Im}\,\zeta}\
\vartheta(z-\zeta). \end{equation} It is holomorphic in $z$ and real analytic in
$\zeta$. It follows that
\begin{equation}A^p_z=-\frac{i}{2}\,\frac{f'}{f}(z;\zeta),\label{gauge-pf}
\end{equation} is meromorphic in $z\in\complex$ with simple poles at $p+m+in$,
each of residue $-i/2$. It obeys
\begin{equation}A^p_z(z+1)=A^p_z(z),\quad A^p_z(z+i)=A^p_z(z)+i\pi,
\end{equation} and is doubly periodic in $p$. It means that the gauge potential
$A^p$ is independent of the choice of $p\in\complex$ modulo the lattice
translations. Clearly, $dA^p=2\pi\delta(z-p)dx\wedge dy$ on the torus $T$, so
Eq.~(\ref{pft}) indeed gives a canonical gauge potential of a unit point flux. 

The periods of 1-form $\dot A$ give the total emf, $V$, about the parallel and
meridian of the torus
$T$. One finds from Eq.~(\ref{gauge-pf}) that $V=E-2\pi i\dot p$. For N point
fluxes with fluxes $\Phi_k$ (each a multiple of $\Phi_0=2\pi$) on the torus with
both parallel and meridian of length $L$ we get Eq.~(\ref{totalemf}) by scaling
and putting back the velocity of light $c$. 

Let us consider now the force equations, Eqs.~(\ref{main}, \ref{forceplane}). By
the principle of virtual work, the {\em quantum observable} associated with the
force on the $j$-th point flux located at $p_j$ is \begin{equation}
F_j=F_{j1}+iF_{j2}=-2\,\frac{\partial H}{\partial\bar p_j}. \label{virtual}
\end{equation} The Landau Hamiltonians we consider are such that for any value
of the parameters $p_j,\ j=0,\dots,N$ (we write $p_0=\phi$), the operator is
unitarily equivalent to the Landau Hamiltonian $H_0=H(A^0)$ {\em without point
fluxes and in the absence of external electric field}, that is, $H(A)=
U(A)H_0U^*(A)$ with an appropriate unitary $U(A)$. (It is here that we use the
fact that the point flux carries an integer flux). Let $P$ denote the spectral
projection on a Landau level of $H(A)$, i.e., $P=\sum_\ell
|\phi_\ell\rangle\langle \phi_\ell|$, with $|\phi_\ell\rangle$ being normalized
eigenstates that span the Landau level. Clearly, $P=U(A)P_0U^*(A)$ with $P_0$
the (fixed) projection on the Landau level of $H_0$. For such a family of
unitarily related Hamiltonians we get from the basic equation of adiabatic
transport \cite{thouless,tknn,seiler91} that \begin{equation}
\langle F_j\rangle= -i\, \sum_{k=0}^N \Omega_{\bar p_j p_k}\, \dot p_k +
O(1/\tau^2),\label{forceequation}\end{equation} where $$\Omega_{\bar p_j
p_k}=Tr\left(P\left[\frac{\partial P} {\partial\bar p_j},\frac{\partial
P}{\partial p_k}\right]P\right)$$ are the components of the trace of the
adiabatic curvature \cite{berry,simon83}: \begin{equation} \Omega
(P)=P\,dP\wedge dP\,P.\label{AC}\end{equation} Using the unitarity of $U$ one
finds, formally, that \begin{equation} Tr\,\Big(\Omega (P)\Big)= Tr\,\Big(P_0
[{\cal A},P_0] \wedge [{\cal A},P_0] P_0\Big),\label{AC'} \end{equation} where
${\cal A}= U^*dU.$ In the case of the torus we use the following formally
equivalent version of the above equation \begin{equation} Tr\,\Big(\Omega
(P)\Big)= Tr\,\Big(P_0 {\cal A}\wedge{\cal A}\Big).\label{ACT} \end{equation}
(The two are strictly equivalent when $P_0 {\cal A}P_0$ is Hilbert-Schmidt.)

To compute the adiabatic curvatures we need to explicitly construct the unitary
families $U(A)$. We shall now outline the construction of $U$ for the case of
the torus. Consider the family of Landau Hamiltonians $H(A)$ for a {\em unit}
point flux given by Eq.~(\ref{landau}), with $$D_A=\frac{\partial}{\partial\bar
z}+\frac{iBy}{2}-\frac{i\phi}{2} +\frac{1}{2}\,\frac{\bar f'}{\bar f}.$$ The
total magnetic flux of $A$ through the unit torus $T$ is $\int_TdA=B+2\pi.$ By
Dirac's quantization condition, $B/2\pi$ is an integer, and we assume that it is
positive. The unitary operators describing magnetic translations which commute
with $D_A$ are \begin{equation} T_1\psi(z)=\psi(z+1),\quad
T_2\psi(z)=e^{i(B+2\pi)x}\,\psi(z+i),\end{equation} that give the usual magnetic
translation boundary conditions \cite{zak}: \begin{equation}T_1\psi=\psi,\quad
T_2\psi=\psi.\label{boundary}\end{equation} For $H_0$ we have, instead, the
boundary conditions \begin{equation}T_1^0\psi=\psi,\quad T_2^0\psi=\psi
\label{b0}\end{equation} with $T_1^0\psi(z)=\psi(z+1),\;
T_2^0\psi(z)=e^{iBx}\,\psi(z+i)$.

We represent
$U$ as a composition of two auxiliary unitary operators $V$ and $W$. The first
one,
\begin{equation} V(t)\psi(z)= e^{-iy\,\mbox{\scriptsize Im}\,t} \psi(z-t/B),
\end{equation} commutes with $T^0_1,\,T^0_2$ and has the property that
\begin{equation} V({i\phi}) D_{A^0} = D_{A^0+A^\phi} V({i\phi}).\label{LAB}
\end{equation} The second one is a multiplication operator  (gauge
transformation)
$$W(p)\psi(z)=e^{\pi(\zeta\bar z-\bar\zeta z)}\;\frac{f}{|f|}(z;\zeta)\,\psi(z),
\quad\zeta=p+\frac{1}{2}+\frac{i}{2}, $$ 
boundary conditions in Eqs.~(\ref{boundary}, \ref{b0}):
$$W(p)T_1^0=T_1W(p),\quad W(p)T_2^0=T_2W(p).$$ Now put $U(A)=W(p)V(t)$ with
$t=i\phi-2\pi\zeta$. We see that \begin{equation}U(A)D_{A^0}=D_AU(A),\quad\quad
A=A^0+A^\phi+A^p. \end{equation} Since the unitary $W(p)$ is a multiplication
operator, the eigenstate bundles for the family $H(A)=U(A)H_0U^*(A)$ are
topologically equivalent to those of $V(t)H_0V(t)^*$ with $t=i\phi-2\pi\zeta$.
If $W(p)$ were a smooth family, the eigenstate adiabatic curvatures would also
coincide, but $W(p)$ is not. This gap can be patched by an appropriate limiting
argument. Substituting the formulas \begin{eqnarray*}
&&V(t)^*dV(t)=-\left(\frac{1}{B}\frac{\partial}{\partial z}+\frac{iy}{2}
\right)dt-\left(\frac{1}{B}\frac{\partial}{\partial\bar z}+\frac{iy}{2}
\right)d\bar t,
\\ &&V(t)^*dV(t)\wedge V(t)^*dV(t)=\frac{1}{2B} dt\wedge d\bar t,\label{V}
\end{eqnarray*} into Eq.~(\ref{ACT}) for the family $P=V(t)P_0V(t)^*$ with
$t=i\phi-2\pi\zeta$, we get \begin{equation}
Tr\,\Omega\Big(V({it})P_0V^*({it})\Big) =-\frac{1}{4\pi} dt\wedge d\bar
t,\label{adiabatic}\end{equation} where $dt=i\,d\phi-2\pi\,dp$. For $N$ point
fluxes $p_k$ with fluxes $\Phi_k$ on the torus associated with the $L\times L$
square, by scaling we again get Eq.~(\ref{adiabatic}) with \begin{equation}
dt=iLd\phi-\frac{1}{L}\,\sum_{k=1}^N \Phi_k\,dp_k.\end{equation} Putting back
the constants $c$ and $\frac{e}{\hbar}$ and using Eq.~(\ref{forceequation}) we
obtain Eq.~(\ref{main})

One can make an independent formal calculation of the forces on a point flux in
the Euclidean plane starting with Eq.~(\ref{AC'}) and making explicit
computations with integral kernels following the methods in
\cite{ass,bellissard}. The integral kernel of $P_0$ is Gaussian, and the unitary
$U$ in the plane is an abelian gauge transformation with $${\cal
A}=\frac{1}{2}\left(\sum -\frac{dp_j}{z-p_j} + \frac{d\bar p_j} {\bar z-\bar
p_j}\right) +iz d\bar\phi +i \bar z d \phi .$$ The resulting multidimensional
(singular) integrals can be evaluated explicitly and give the same result that
one obtains from the torus calculation by setting $L\to\infty$. Note that one
can not use Eq.~(\ref{ACT}) (which gives zero since $\cal A$ is abelian) because
$P_0$ is infinite dimensional, and $P_0{\cal A}P_0$ is not Hilbert-Schmidt. 

The effective electric charging of point flux is a way of describing the force
that a point flux experiences in an external electric field. This raises the
question: is this charging real? For the sake of simplicity and concreteness let
us focus on the Euclidean plane. The charge associated with a point flux is a
subtle issue in the sense that one gets different answers depending on the
precise question one asks. It is a basic fact about integer quantum Hall systems
that adding a point flux sends an integral number of charges to infinity
\cite{laughlin}. In this sense there is a real charge associated with the point
flux. However, an {\em a priori} different notion of charge associated with a
point flux is to look at the charge density in the neighborhood of the flux. In
Chern-Simons theory of the quantum Hall effect \cite{stone} one has, indeed, an
excess charge density near a point flux. On the other hand, for an {\em ideal
point flux} which carries an integral number of flux quanta, it is a consequence
of gauge invariance that the charge distribution of a full Landau level in the
plane (both for Pauli and Schr\"odinger Hamiltonian) is blind to presence of the
point flux and in this sense no excess charge is attached to a point flux.

The forces that act on point fluxes display neither symmetry nor duality between
point charges and point fluxes and appear to conflict with Galilei invariance.
This is most dramatic in the case of the Euclidean plane where there is a
non-vanishing electric force but no Lorentz force. In the case of the torus one
finds both an electric and Lorentz type force, but the two are not related in
the way one would expect from Lorentz and Galilei invariance. 

In the case of the torus one has a simple reason to doubt arguments based on
Lorentz invariance because Lorentz invariance is broken by the identifications
that make a torus out of a square. In the case of the plane it is less obvious
what has gone awry with Lorentz invariance. The answer to this is that taking
the adiabatic limit breaks Lorentz and Galilei invariance. The adiabatic setting
stipulates that in the distant past the Hamiltonian is a {\em time independent}
Landau Hamiltonian and the initial state of the system is associated with a
spectral projection and a full Landau level. This distinguishes a frame and
breaks Galilei invariance.

\section*{Acknowledgments}

This work was partially supported by a grant from the Israel Academy of
Sciences, the Deutsche Forschungsgemeinschaft, and by the Fund for Promotion of
Research at the Technion. In addition to that, the work of PZ was supported in
part by a grant from the Russian Fund for Fundamental Research and by the
Rosenbaum Foundation, and he also acknowledges the hospitality of ITP at the
Technion where the first stage of this work was done.

\end{document}